Laboratoire National Saturne
CEN Saclay
91191 Gif Sur Yvette Cedex
France

...

...

...

# $\pi^-$ tagging in $\Lambda \rightarrow p\pi$ decay in homogeneous and inhomogeneous magnetic fields.

Luca Cheli
*1n2p3 LNS Saclay*

March 17, 1995

# 1 Introduction.

Our research was targeted to analyze, through computer simulations, the resolution power of a spectrometer with different kinds of magnetic fields and at different energies.

Throughout our analysis, we have focused our attention on the $\pi^-$ tagging in the $\Lambda \to p\pi$ reaction.

We wanted to study the possibility of making a good tagging at high energies ($\geq 100 GeV$), i.e. a good $\Lambda$ momentum reconstruction with the software at our disposition.

To this end the work has been divided in two parts:

- The experimental setup definition.
- The proper simulation.

# 2 Definition of the setup.

In line of principle the most straightforward definition of an experimental setup of this kind must concentrate on three points:

1. The target.
2. The magnet (i.e. the magnetic field).
3. The detectors.

## 2.1 The target.

At our level of analysis the presence of a real, material target was deeemed unnecessary, so that, to simplify the study and not to introduce in it too many variables, we have decided to identify it as a geometric point in a first time and with an area later.

## 2.2 The magnet and the field.

To greatly simplify the first part of the analysis and to be able to use quick and simple analytic formulas for the motion of particles in a magnetic field, we have decided to define initially the field as homogeneous of dipolar type with:

$$B_x = 0 , \; B_y = 0 , \; B_z = const$$

The constant value of the field has been set to 1.5 or 3 T, according to the momentum of the $\Lambda$.

The study of the effect of more complex and more interesting field configuration has been delayed to a later time, when the problem had been correctly mastered in its easier configuration.

## 2.3 The detectors.

The detectors' dimensionment and placing was certainly the more complicated task of the first part of this study, because their optimization depends closely on the chosen magnetic field, and we wanted to modify this field during the study. At the same time there is also an energy dependence of the positionment and of the resolution: at different value of the $\Lambda$ momentum ($p_\Lambda$), different resolution are required to obtain the same precision in the momentum reconstruction and different detector positions are required to take in account the different mean life of the $\Lambda$.

To this end and because of the necessities of our momentum reconstruction program, we have decided to put an high number of detector of the same standardized form. This form is next to the optimum only for the dipolar field, but we think that the generality of our analysis can go along with the lack of some optimization.

The positionment of the detectors was made considering the particle motion in the uniform field at different momenta in a range from 0.5 to more than 100 GeV/c.

Such a wide range forced us to a compromise among the best possibilities present at lower ($\leq 10 Gev/c$) and higher momenta. The final setup is intended to perform decently for momenta up to 250 GeV/c, but clearly is not the most efficient or cheaper that can be envisaged for a particular momentum subrange or a different kind of field.

This setup is showed in 1 within the reference system used for the simulations.

The detectors 1-3 have a resolution of $250 \mu m$, the detectors 5 - 8 of $0.5 cm$ and the others of $1 cm$.

The detector distances from the target (origin) and their angular openings are summarized in Table 1.

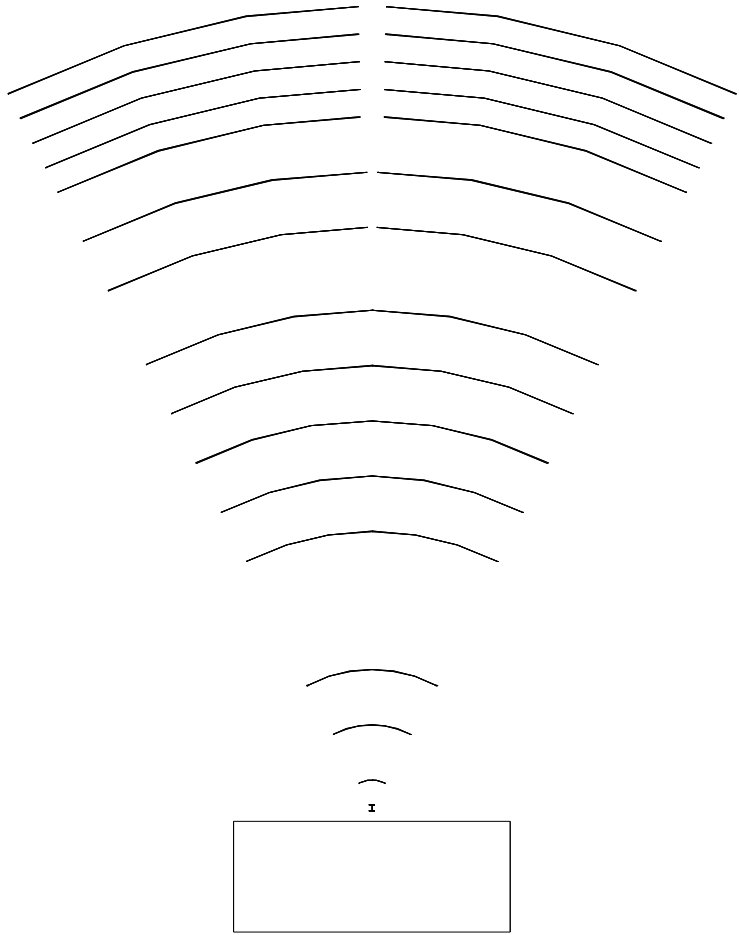

Figure 1: Detector position in the reference system of the simulations. The dummy target is the small square. The object before the target is a catcher intended to stop low momentum particles from spiralizing around the target.

| TABLE 1 - Detectors positions and openings. | | |
| --- | --- | --- |
| DETECTOR NUMBER | RADIUS (cm) | $\phi$ (degrees) |
| 1 | 10 | 2 * 28 |
| 2 | 30 | 2 * 28 |
| 3 | 50 | 2 * 28 |
| 4 | 100 | 2 * 27 |
| 5 | 120 | 2 * 27 |
| 6 | 140 | 2 * 27 |
| 7 | 160 | 2 * 27 |
| 8 | 180 | 2 * 27 |
| 9 | 210 | 2 * 26.5 |
| 10 | 230 | 2 * 26.5 |
| 11 | 250 | 2 * 26 |
| 12 | 260 | 2 * 26 |
| 13 | 270 | 2 * 26 |
| 14 | 280 | 2 * 26 |
| 15 | 290 | 2 * 26 |

As illustrated in 1, a 0.5 - 1 degree open window has been left in the more distant detectors to allow very high momenta to pass through without adding to the background.

For the purposes of definition in the simulation software, the thickness of the detectors has been specified in 1 mm, but this hasn' t a big relief, because the detectors are considered as if they were made of vacuum, that is not interacting. This simplification was done in order to be able to evaluate the effects of the different fields without superimposing a random scattering component.

## 3 The simulation.

For the computer simulation we have used mainly two programs: a Monte Carlo simulation program based on the GEANT package [1] and a momentum reconstruction program essentially based on the MOMENTM routine [3], which was included up to some time ago in the CERN libraries.

The physical problem on which we focused our efforts was the determination of the momentum of a decaying $\Lambda$ particle through the detection of the generated pion alone.

This technique (" tagging ") has the advantage of allowing one to deal with lower momenta, in fact the pion carries away at most approximately one fourth of the momentum of the decaying $\Lambda$. So the momentum reconstruction becomes easier because the magnetic rigidity is reduced in the same ratio and the angle of emission at the decay increases.

### 3.1 Kinematics.

In a two body reaction we can write [5], in the lab reference system:

$$2p_1 p_3 cos\theta_3 = m_4^2 - m_1^2 - m_2^2 - m_3^2 + 2(E_1 + m_2)E_3 - 2E_1 m_2$$

in our case ($\Lambda$ decay) we have that subscript 1 refers to the $\Lambda$, particle 2 is non existent, 3 is referred to the pion and 4 to the proton, so that we can simplify to obtain:

$$2p_\Lambda p_\pi cos\theta_{\lambda\pi} = m_p^2 - m_\Lambda^2 - m_\pi^2 + 2E_\Lambda E_\pi$$

if we remember that :

$$cos\theta_{\Lambda\pi} = \frac{p_{\pi x}}{p_\pi}$$

if we assume that $\vec{p_\Lambda}$ is parallel to the $x$ axis, with a bit of algebra we can obtain for the $\Lambda$ momentum the following expression:

$$p_\Lambda = \frac{-V p_{\pi x} \pm E_\pi \sqrt{V^2 + m_\Lambda^2 (p_{\pi x}^2 - E_\pi^2)}}{E_\pi^2 - p_{\pi x^2}} \qquad (1)$$

where

$V = \frac{m_p^2 - m_\Lambda^2 - m_\pi^2}{2}$ .

As the $\theta_{\Lambda\pi}$ angle becomes smaller with the growth of $p_\Lambda$, the reconstruction of the $p_\pi$ x component becomes more and more critical : a small error can lead the square root in equation 1 to become imaginary.

Other formulas have been tried, but they behave in no better, or even worse manner.

### 3.2 Structure of the simulation.

The Monte Carlo simulation program is based on GEANT, it generates a $\Lambda$ particle in a predetermined zone of space, generates the decay, then tracks it according to the laws of kinematics and of motion of a charged particle in a given magnetic field.

Hits of the particles with the simulated detectors are registered. This recorded hits are fed as input in the momentum reconstruction program, which outputs the momentum module and components of the concerned particle. This particle, in our tagging, is the pion, knowledge of these data allows us to reconstruct $p_\Lambda$ through the 1.

The MOMENTM routine, which is the core of our momentum recostruction program, gives its output (momenta) in the position corresponding to the first space point in input. We of course do not know the decay vertex position, because it is ruled by randomness through the low of particle decay, but we must have the $x$ component of the pion momentum in the vertex with the highest possible precision, because the formula 1 is very sensitive to this valule, especially as the energy grows and the decay angle decreases. So we have made a modification in the original MOMENTM routine in order to be able to find the decay point, unfortunately the method employed is rigorously valid only if the $\lambda$ momentum vector is perfectly coincident with one of the axes of our reference system. This condition is easily satisfied if the $\Lambda$ is generated in the origin with the required direction of the momentum vector, but otherwise it is violated, so we will able to reconstruct with the best accuracy only in this particular case. If the quoted condition is not satisfied, the vertex finding method still gives meaningful approximate results, but they deteriorate quickly the more we are far from the best situation, and above all with the rising in energy.

## 4 The results.

We have done our simulation with three kind of magnetic field, whose analytic expressions are :

$$B_x = 0 \ , \ B_y = 0 \ , \ B_z = const \tag{2}$$

for the homogeneous field,

$$B_x = 0 \ , \ B_y = 0 \ , \ B_z = const \ \times \ atan(\frac{x}{y}) \tag{3}$$

and

$$B_x = 0 \ , \ B_y = B_z = \frac{const}{r + 20} \tag{4}$$

where

$$r \ = \ \frac{\sqrt{y^2 + z^2}}{10}$$

if the coordinates are in centimeters and the field in kilogauss (GEANT units).

We made two main sets of simulations, one with $p_\Lambda \ = \ 5 \ GeV/c$ and the other with $p_\Lambda \ = \ 250 \ GeV/c$, the constants were calculated to give $B_{max} \ = \ 1.5T$ in the first case and $B_{max} \ = \ 3.0T$ in the second.

In both cases we made separate simulations for $\Lambda$ generated in and outside the origin of the reference system, always with $\vec{p_\Lambda}$ parallel to the $x$ axis.

### 4.1 Runs at 5 GeV/c.

With the field 2 we obtained $\frac{\Delta p_\Lambda}{p_\Lambda} \ \approx \ 0.1$ % both for particles starting in the origin or a few centimeters outside.

With the field 3 we had $\frac{\Delta p_\Lambda}{p_\Lambda} \ \approx \ 0.3$ % again for both kinds of $\Lambda$ generation.

With the field 4 the results were $\frac{\Delta p_\Lambda}{p_\Lambda} \ \approx \ 10$ % for a $\Lambda$ starting $1.5cm$ out of the origin and $\frac{\Delta p_\Lambda}{p_\Lambda} \ \approx \ 3$ % for the well behaved case.

### 4.2 Runs at 250 GeV/c

Equation 1 is now much more critical and we had $\frac{\Delta p_\Lambda}{p_\Lambda} \ \approx \ 3.5$ % for the field 2 and the $\Lambda$ generated in the origin.

If the the generation occurs at $r \ = \ 0.15cm$ we still have $\frac{\Delta p_\Lambda}{p_\Lambda} \ \approx \ 4$ %, but if $r \ = \ 0.3cm$, the square root in 1 becomes immaginary and our formula fails.

With the fields 3 and 4 we had $\frac{\Delta p_\Lambda}{p_\Lambda} \ \approx \ 35$ % and $\frac{\Delta p_\Lambda}{p_\Lambda} \ \approx \ 75$ % respectively, already in the case of the $\Lambda$ starting from the origin.

## 5 Conclusion.

Our simulations have shown that the tagging is a usable and useful tecnique at momentum range of a few GeV/c even in a inhomogeneous field. On the other side it is very difficult to apply beyond a few tens of GeV/c.

Actually we made a few simulations for $p_\Lambda = 20 GeV/c$ and for $p_\Lambda = 100 GeV/c$. In the first case the resolution on $p_\Lambda$ worsened of a factor of at least two, with respect to the $5 GeV/c$ runs, while in the second case the situation was almost identical to that of the $250 GeV/c$ runs.

So we can conclude that without a more refined and specialized software for the momentum reconstruction (and perhaps refined and specialized detector setup) the tecnique illustrated for this study appears better suited for low and intermediate energies experiments.